\documentclass[aps,prd,superscriptaddress,showpacs,nofootinbib,floatfix,twocolumn]{revtex4}

\usepackage{graphicx}	

\begin{document}


\title{Inclusive cross section and double helicity asymmetry \\
for $\pi^{0}$ production in $p+p$ collisions at $\sqrt{s} = 62.4$ GeV}

\newcommand{\abilene}{Abilene Christian University, Abilene, TX 79699, U.S.}
\newcommand{\acadsin}{Institute of Physics, Academia Sinica, Taipei 11529, Taiwan}
\newcommand{\banaras}{Department of Physics, Banaras Hindu University, Varanasi 221005, India}
\newcommand{\barc}{Bhabha Atomic Research Centre, Bombay 400 085, India}
\newcommand{\bnlchem}{Chemistry Department, Brookhaven National Laboratory, Upton, NY 11973-5000, U.S.}
\newcommand{\bnlcoll}{Collider-Accelerator Department, Brookhaven National Laboratory, Upton, NY 11973-5000, U.S.}
\newcommand{\bnlphys}{Physics Department, Brookhaven National Laboratory, Upton, NY 11973-5000, U.S.}
\newcommand{\caucr}{University of California - Riverside, Riverside, CA 92521, U.S.}
\newcommand{\charlesczech}{Charles University, Ovocn\'{y} trh 5, Praha 1, 116 36, Prague, Czech Republic}
\newcommand{\ciae}{China Institute of Atomic Energy (CIAE), Beijing, People's Republic of China}
\newcommand{\cns}{Center for Nuclear Study, Graduate School of Science, University of Tokyo, 7-3-1 Hongo, Bunkyo, Tokyo 113-0033, Japan}
\newcommand{\colorado}{University of Colorado, Boulder, CO 80309, U.S.}
\newcommand{\columbia}{Columbia University, New York, NY 10027 and Nevis Laboratories, Irvington, NY 10533, U.S.}
\newcommand{\czechtech}{Czech Technical University, Zikova 4, 166 36 Prague 6, Czech Republic}
\newcommand{\dapnia}{Dapnia, CEA Saclay, F-91191, Gif-sur-Yvette, France}
\newcommand{\debrecen}{Debrecen University, H-4010 Debrecen, Egyetem t{\'e}r 1, Hungary}
\newcommand{\elte}{ELTE, E{\"o}tv{\"o}s Lor{\'a}nd University, H - 1117 Budapest, P{\'a}zm{\'a}ny P. s. 1/A, Hungary}
\newcommand{\fit}{Florida Institute of Technology, Melbourne, FL 32901, U.S.}
\newcommand{\fsu}{Florida State University, Tallahassee, FL 32306, U.S.}
\newcommand{\gsu}{Georgia State University, Atlanta, GA 30303, U.S.}
\newcommand{\hiroshima}{Hiroshima University, Kagamiyama, Higashi-Hiroshima 739-8526, Japan}
\newcommand{\ihepprot}{IHEP Protvino, State Research Center of Russian Federation, Institute for High Energy Physics, Protvino, 142281, Russia}
\newcommand{\illuiuc}{University of Illinois at Urbana-Champaign, Urbana, IL 61801, U.S.}
\newcommand{\instpasczech}{Institute of Physics, Academy of Sciences of the Czech Republic, Na Slovance 2, 182 21 Prague 8, Czech Republic}
\newcommand{\isu}{Iowa State University, Ames, IA 50011, U.S.}
\newcommand{\jinrdubna}{Joint Institute for Nuclear Research, 141980 Dubna, Moscow Region, Russia}
\newcommand{\kek}{KEK, High Energy Accelerator Research Organization, Tsukuba, Ibaraki 305-0801, Japan}
\newcommand{\kfki}{KFKI Research Institute for Particle and Nuclear Physics of the Hungarian Academy of Sciences (MTA KFKI RMKI), H-1525 Budapest 114, POBox 49, Budapest, Hungary}
\newcommand{\korea}{Korea University, Seoul, 136-701, Korea}
\newcommand{\kurchatov}{Russian Research Center ``Kurchatov Institute", Moscow, Russia}
\newcommand{\kyoto}{Kyoto University, Kyoto 606-8502, Japan}
\newcommand{\labllr}{Laboratoire Leprince-Ringuet, Ecole Polytechnique, CNRS-IN2P3, Route de Saclay, F-91128, Palaiseau, France}
\newcommand{\lawllnl}{Lawrence Livermore National Laboratory, Livermore, CA 94550, U.S.}
\newcommand{\losalamos}{Los Alamos National Laboratory, Los Alamos, NM 87545, U.S.}
\newcommand{\lpc}{LPC, Universit{\'e} Blaise Pascal, CNRS-IN2P3, Clermont-Fd, 63177 Aubiere Cedex, France}
\newcommand{\lund}{Department of Physics, Lund University, Box 118, SE-221 00 Lund, Sweden}
\newcommand{\mass}{Department of Physics, University of Massachusetts, Amherst, MA 01003-9337, U.S. }
\newcommand{\muenster}{Institut f\"ur Kernphysik, University of Muenster, D-48149 Muenster, Germany}
\newcommand{\muhlenberg}{Muhlenberg University, Allentown, PA 18104-5586, U.S.}
\newcommand{\myongji}{Myongji University, Yongin, Kyonggido 449-728, Korea}
\newcommand{\nagasaki}{Nagasaki Institute of Applied Science, Nagasaki-shi, Nagasaki 851-0193, Japan}
\newcommand{\newmex}{University of New Mexico, Albuquerque, NM 87131, U.S. }
\newcommand{\nmsu}{New Mexico State University, Las Cruces, NM 88003, U.S.}
\newcommand{\ornl}{Oak Ridge National Laboratory, Oak Ridge, TN 37831, U.S.}
\newcommand{\orsay}{IPN-Orsay, Universite Paris Sud, CNRS-IN2P3, BP1, F-91406, Orsay, France}
\newcommand{\peking}{Peking University, Beijing, People's Republic of China}
\newcommand{\pnpi}{PNPI, Petersburg Nuclear Physics Institute, Gatchina, Leningrad region, 188300, Russia}
\newcommand{\riken}{RIKEN, The Institute of Physical and Chemical Research, Wako, Saitama 351-0198, Japan}
\newcommand{\rikjrbrc}{RIKEN BNL Research Center, Brookhaven National Laboratory, Upton, NY 11973-5000, U.S.}
\newcommand{\rikkyo}{Physics Department, Rikkyo University, 3-34-1 Nishi-Ikebukuro, Toshima, Tokyo 171-8501, Japan}
\newcommand{\saispbstu}{Saint Petersburg State Polytechnic University, St. Petersburg, Russia}
\newcommand{\saopaulo}{Universidade de S{\~a}o Paulo, Instituto de F\'{\i}sica, Caixa Postal 66318, S{\~a}o Paulo CEP05315-970, Brazil}
\newcommand{\seoulnat}{System Electronics Laboratory, Seoul National University, Seoul, Korea}
\newcommand{\stonybrkc}{Chemistry Department, Stony Brook University, Stony Brook, SUNY, NY 11794-3400, U.S.}
\newcommand{\stonycrkp}{Department of Physics and Astronomy, Stony Brook University, SUNY, Stony Brook, NY 11794, U.S.}
\newcommand{\subatech}{SUBATECH (Ecole des Mines de Nantes, CNRS-IN2P3, Universit{\'e} de Nantes) BP 20722 - 44307, Nantes, France}
\newcommand{\tenn}{University of Tennessee, Knoxville, TN 37996, U.S.}
\newcommand{\titech}{Department of Physics, Tokyo Institute of Technology, Oh-okayama, Meguro, Tokyo 152-8551, Japan}
\newcommand{\tsukuba}{Institute of Physics, University of Tsukuba, Tsukuba, Ibaraki 305, Japan}
\newcommand{\vandy}{Vanderbilt University, Nashville, TN 37235, U.S.}
\newcommand{\waseda}{Waseda University, Advanced Research Institute for Science and Engineering, 17 Kikui-cho, Shinjuku-ku, Tokyo 162-0044, Japan}
\newcommand{\weizmann}{Weizmann Institute, Rehovot 76100, Israel}
\newcommand{\yonsei}{Yonsei University, IPAP, Seoul 120-749, Korea}
\affiliation{\abilene}
\affiliation{\acadsin}
\affiliation{\banaras}
\affiliation{\barc}
\affiliation{\bnlchem}
\affiliation{\bnlcoll}
\affiliation{\bnlphys}
\affiliation{\caucr}
\affiliation{\charlesczech}
\affiliation{\ciae}
\affiliation{\cns}
\affiliation{\colorado}
\affiliation{\columbia}
\affiliation{\czechtech}
\affiliation{\dapnia}
\affiliation{\debrecen}
\affiliation{\elte}
\affiliation{\fit}
\affiliation{\fsu}
\affiliation{\gsu}
\affiliation{\hiroshima}
\affiliation{\ihepprot}
\affiliation{\illuiuc}
\affiliation{\instpasczech}
\affiliation{\isu}
\affiliation{\jinrdubna}
\affiliation{\kek}
\affiliation{\kfki}
\affiliation{\korea}
\affiliation{\kurchatov}
\affiliation{\kyoto}
\affiliation{\labllr}
\affiliation{\lawllnl}
\affiliation{\losalamos}
\affiliation{\lpc}
\affiliation{\lund}
\affiliation{\mass}
\affiliation{\muenster}
\affiliation{\muhlenberg}
\affiliation{\myongji}
\affiliation{\nagasaki}
\affiliation{\newmex}
\affiliation{\nmsu}
\affiliation{\ornl}
\affiliation{\orsay}
\affiliation{\peking}
\affiliation{\pnpi}
\affiliation{\riken}
\affiliation{\rikjrbrc}
\affiliation{\rikkyo}
\affiliation{\saispbstu}
\affiliation{\saopaulo}
\affiliation{\seoulnat}
\affiliation{\stonybrkc}
\affiliation{\stonycrkp}
\affiliation{\subatech}
\affiliation{\tenn}
\affiliation{\titech}
\affiliation{\tsukuba}
\affiliation{\vandy}
\affiliation{\waseda}
\affiliation{\weizmann}
\affiliation{\yonsei}
\author{A.~Adare}	\affiliation{\colorado}
\author{S.~Afanasiev}	\affiliation{\jinrdubna}
\author{C.~Aidala}	\affiliation{\mass}
\author{N.N.~Ajitanand}	\affiliation{\stonybrkc}
\author{Y.~Akiba}	\affiliation{\riken} \affiliation{\rikjrbrc}
\author{H.~Al-Bataineh}	\affiliation{\nmsu}
\author{J.~Alexander}	\affiliation{\stonybrkc}
\author{K.~Aoki}	\affiliation{\kyoto} \affiliation{\riken}
\author{L.~Aphecetche}	\affiliation{\subatech}
\author{J.~Asai}	\affiliation{\riken}
% ...
\collaboration{PHENIX Collaboration} \noaffiliation

\date{\today}

\begin{abstract}

The PHENIX experiment presents results from the RHIC 2006 run with 
polarized $p+p$ collisions at $\sqrt{s}=62.4$ GeV, for inclusive 
$\pi^{0}$ production at mid-rapidity.  Unpolarized cross section 
results are measured for transverse momenta $p_T=0.5$ to $7$ GeV/$c$.  
Next-to-leading order perturbative quantum chromodynamics calculations 
are compared with the data, and while the calculations are consistent 
with the measurements, next-to-leading logarithmic corrections improve 
the agreement. Double helicity asymmetries $A_{LL}$ are presented for 
$p_T=1$ to $4$ GeV/$c$ and probe the higher range of Bjorken $x$ of 
the gluon ($x_g$) with better statistical precision than our previous 
measurements at $\sqrt{s}=200$ GeV. These measurements are sensitive 
to the gluon polarization in the proton for $0.06<x_g<0.4$.

\end{abstract}

	
\maketitle

\section{Introduction}

Spin is a property of particles as fundamental as charge and mass. The 
spin of the proton was first determined in the 1920s, yet we still do 
not have a detailed understanding of what inside the proton makes up 
the spin of the proton. Polarized lepton-nucleon deep-inelastic 
scattering (DIS) experiments have revealed that only $\sim$25\% of the 
proton spin can be attributed to the spins of the quarks and 
anti-quarks \cite{EMC,sDIS} indicating that the proton spin must be 
largely carried by the spin of the gluons and/or orbital angular 
momentum of quarks and gluons. Polarized proton-proton collisions at 
the Relativistic Heavy-Ion Collider (RHIC) provide a laboratory to 
study the gluon-spin contribution to the proton spin structure, 
$\Delta G$, with strongly interacting probes via measurements of 
double helicity asymmetries ($A_{LL}$) \cite{rhic_spin}.

The $A_{LL}$ of $\pi^0$'s is defined as
\begin{equation}
  A_{LL}^{\pi^0} = \frac{\sigma_{++}-\sigma_{+-}}{\sigma_{++}+\sigma_{+-}},
\label{eq:a_ll_sigma}
\end{equation}
where $\sigma_{++}(\sigma_{+-})$ represents the $\pi^0$ production cross 
section in polarized $p+p$ collisions with the same (opposite) 
helicities. In leading order (LO) perturbative Quantum Chromodynamics 
(pQCD), $\pi^0$ production is the sum of all possible subprocesses $ab 
\rightarrow cX$, where $a$,$b$ represent the initial partons in the 
protons, $c$ is the final state parton which fragments into a $\pi^0$, 
and $X$ is the unobserved parton. Then $A_{LL}$ is calculated as
\begin{equation}
A_{LL}^{\pi^0} = \frac{ \Sigma_{a,b,c} \Delta f_a \Delta f_b
\hat{\sigma}^{[a b \rightarrow cX]} \hat{a}_{LL}^{[ab\rightarrow cX]}D_c^{\pi^0}}
{\Sigma_{a,b,c} {f_a} {f_b} \hat{\sigma}^{[ab \rightarrow cX]} D_c^{\pi^0}},
\end{equation}
where $f_{a,b}$ represent unpolarized parton distribution functions 
(PDF) of parton $a,b$ and $\Delta f_{a,b}$ represent polarized PDFs, 
$D_c^{\pi^0}$ is a fragmentation function (FF) of parton $c$ to $\pi^0$, 
$\hat{\sigma}^{[ab \rightarrow cX]}$ and $\hat{a}_{LL}^{[a b \rightarrow 
cX]}$ denote respectively the cross section and $A_{LL}$ of the 
subprocess $ab \rightarrow cX$. The sum is performed for all possible 
partons (quarks and gluons). The Bjorken-$x$ dependence of the PDFs, the 
kinematical dependence of the FFs, and the integral over all possible 
kinematics are omitted in the equation. The partonic quantities 
$\hat{\sigma}$ and $\hat{a}_{LL}$ can be calculated in pQCD. Since 
$\pi^0$ production is dominated by gluon-gluon and quark-gluon 
scattering in the measured $p_T$ range ($p_T < 4$~GeV/$c$), $A_{LL}$ is 
directly sensitive to the polarized gluon distribution function in the 
proton.

Cross section measurements at RHIC have established the validity of 
using a next-to-leading order (NLO) pQCD description at $\sqrt{s}=200$ 
GeV for inclusive mid-rapidity $\pi^{0}$ 
\cite{pi0cross_run2,pi0cross_all_run5} and forward $\pi^{0}$ 
production~\cite{pi0_star}, and for mid-rapidity jet~\cite{jet_star} 
and direct photon production~\cite{photon_phenix}. However, at lower 
center of mass energy, NLO pQCD calculations have been less successful 
in describing the data~\cite{low_energy}.  The inclusion of 
``threshold resummation'' at next-to-leading logarithmic accuracy 
(NLL)~\cite{nll} improves the agreement between theory and data at 
fixed-target energies. While taking into account threshold logarithms 
at the fixed-target kinematic region is essential, they may also need 
to be accounted for at $\sqrt{s}=62.4$ GeV, but will provide a smaller 
effect~\cite{nlo_nll}.

A precise measurement of the inclusive $\pi^0$ production cross 
section at $\sqrt{s}=62.4$~GeV is important for the heavy-ion program 
at RHIC. A new state of dense matter is formed in Au+Au collisions at 
200~GeV and parton energy loss in the produced dense medium results in 
high $p_T$ leading hadron suppression. Measurements of high $p_T$ data 
at lower energies are of great importance in identifying the energy range 
at which the suppression sets in. However, they require solid 
measurements of the cross section in $p+p$ collisions as a baseline 
for medium effects. At the ISR, inclusive neutral and charged pion 
cross sections were measured several times at $\sqrt{s} \sim 
62$~GeV~\cite{ISR_n,ISR_pi}, but they have large uncertainties and 
have a large variation~\cite{david_ISR}. Having both heavy-ion and 
baseline $p+p$ measurements with the same experiment is advantageous 
as it leads to a reduction of the systematic uncertainties and, thus, 
to a more precise relative comparison of the data.

In this paper, we present results on inclusive neutral pion production 
at mid-rapidity from proton-proton collisions at $\sqrt{s}=62.4$ GeV 
from data collected during the RHIC 2006 run. A sample of events from 
longitudinally polarized $p+p$ collisions (about 2/3 of the total data 
sample) was used for double helicity asymmetry measurements. The other 
events from the 2006 data sample were from transversely polarized 
$p+p$ collisions and, along with the longitudinally polarized data, 
were used for the unpolarized cross section measurements, by averaging 
over the different initial spin states.

The structure of this paper is as follows. The PHENIX subsystems used 
in this analysis are briefly introduced in section 
\ref{label:experiment}. The unpolarized $\pi^0$ cross section analysis 
and the results are discussed in section \ref{label:xsec}. The $\pi^0$ 
$A_{LL}$ analysis and the results follow in section \ref{label:all}, 
and a summary is given in section \ref{label:summary}.

\section{EXPERIMENT}
\label{label:experiment}

The PHENIX experiment at RHIC measured $\pi^0$'s via $\pi^0 
\rightarrow \gamma\gamma$ decays using a highly segmented ($\Delta 
\eta \times \Delta \phi \sim 0.01 \times 0.01$) electromagnetic 
calorimeter (EMCal)~\cite{nim_emc}, covering a pseudorapidity range of 
$|\eta| < 0.35$ and azimuthal angle range of $\Delta \phi = \pi$. The 
EMCal comprises two calorimeter types: 6 sectors of lead scintillator 
sampling calorimeter (PbSc) and 2 sectors of a lead glass Cherenkov 
calorimeter (PbGl). Each of the EMCal towers was calibrated by the 
two-photon invariant mass from $\pi^0$ decays and cross checked 
against the energy deposited by the minimum ionizing particles in the 
EMCal, and the correlation between the measured momenta of electron 
and positron tracks and the associated energy deposited in the EMCal. 
The uncertainty on the absolute energy scale was 1.2\%.

The $\pi^0$ data in this analysis were collected using two different 
triggers. One is a beam-beam counter (BBC) trigger which was defined 
by the coincidence of signals in two BBCs located at pseudorapidities 
$\pm (3.0-3.9)$ with full azimuthal coverage~\cite{nim_bbc}. The time 
difference between the two BBCs was used to determine the collision 
vertex along the beam axis, which in this analysis was required to be 
within 30 cm from the center of the PHENIX interaction region (IR). 
The other trigger is an EMCal-based high $p_T$ photon trigger, in 
which threshold discrimination corresponding to a deposited energy of 
~0.8 GeV was applied independently to sums of analog signals from $2 
\times 2$ groupings of adjacent EMCal towers \cite{pi0cross_run2}. 
This trigger had limited efficiency for $\pi^0$ detection at low $p_T$ 
(e.g. ~50\% in 1.0--1.5 GeV/$c$ $p_T$ bin) and close to $100\%$ 
efficiency at $p_T>$3 GeV/$c$.

Beam-beam counters along with zero degree calorimeters 
(ZDC)~\cite{nim_zdc}, which detect neutral particles near the beam 
axis ($\theta<2.5$ mrad), were utilized to determine the integrated 
luminosity for the analyzed data sample needed for the absolute 
normalization of the measured cross sections. Trigger counts defined 
with the BBCs and ZDCs were also used for the precise measurements of 
the relative luminosity between bunches with different spin 
configuration, and the spin dependence of very forward neutron 
production ~\cite{locpol,pi0all_run3,pi0cross_all_run5}, detected by 
the ZDCs, served for monitoring the orientation of the beam 
polarization in the PHENIX interaction region (IR) through the run. 
These are necessary components of the spin asymmetry measurements.

\section{The $pp\rightarrow\pi^0 X$ CROSS SECTION}
\label{label:xsec}

The unpolarized cross section analysis technique was very similar to 
our analyses of $\sqrt{s}=200$ GeV data 
\cite{pi0cross_run2,pi0cross_all_run5} and is briefly discussed in 
section \ref{label:xsec_analysis}. Cross section measurements require 
an absolute determination of luminosity which is described in section 
\ref{label:vernier}. The $\pi^0$ cross section results are presented 
and discussed in section \ref{label:xsec_results}.

\subsection{$\pi^0$ analysis}
\label{label:xsec_analysis}

The $\pi^0$ yield in each $p_T$ bin was determined from the two-photon 
invariant mass spectra. The background contribution under the $\pi^0$ 
peak in the two-photon invariant mass distribution varied from $~75\%$ 
in the lowest 0.5--0.75 GeV/$c$ $p_T$ bin to less than $4\%$ for 
$p_T>3$ GeV/$c$.

\begin{figure}[tbh]
\includegraphics[width=1.0\linewidth]{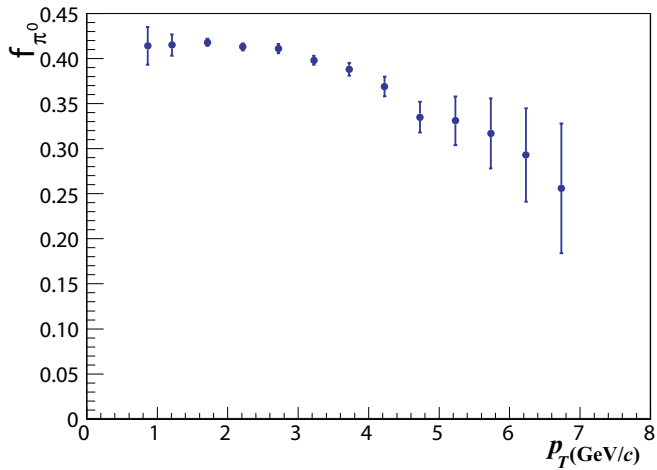}
\caption{\label{fig:trig_bias} (color online)
The fraction of the inclusive $\pi^0$ yield 
which satisfied the BBC trigger condition.}
\end{figure}

One of the main corrections applied to the measured $\pi^0$ spectrum 
is the BBC trigger bias $f_{\pi^0}$, which is defined as the fraction 
of high $p_T$ $\pi^0$ events in the mid-rapidity spectrometer 
acceptance which fire the BBC trigger. This fraction was determined 
from the ratio of the number of reconstructed $\pi^0$ in the high 
$p_T$ photon triggered sample with and without the BBC trigger 
requirement. As shown in Fig.~\ref{fig:trig_bias}, $f_{\pi^0}$ was 
about 40\% up to $p_T\sim3$ GeV/$c$ and then monotonically dropped 
down to ~25\% at $p_T\sim7$ GeV/$c$. The drop can be explained by the 
fact that most of the energy is used for the production of high energy 
jets which contain the measured high $p_T$ $\pi^0$ and there is not 
enough energy left to produce particles in the BBC acceptance $3.0 
\leq |\eta| \leq 3.9$, which was optimized for $\sqrt{s}=200$~GeV 
(where such a drop was not observed \cite{pi0cross_run2}) and was not 
moved for the present $\sqrt{s}=62.4$~GeV measurements.

The main contributors to the systematic uncertainties of the measured 
$\pi^0$ spectrum are given in Table~\ref{tab:errors}. The ``Energy 
scale'' uncertainty includes uncertainties due to EMCal energy 
absolute calibration and nonlinearity. The ``Yield extraction'' 
uncertainty comes from the background subtraction. The ``Yield 
correction'' uncertainty comes from the correction for the geometric 
acceptance, trigger efficiency, reconstruction efficiencies, detector 
response, and photon conversion. The normalization uncertainty is not 
included and is discussed in section \ref{label:vernier}.

\begin{table}[htb]
\caption{Main systematic uncertainties in \% of the $\pi^0$ spectrum 
from the PbSc for two representative $p_T$ bins (the PbGl 
uncertainties are similar).}
\label{tab:errors}
\begin{ruledtabular} \begin{tabular}{lcc} 
$\langle p_T \rangle$ (GeV/$c$) & 1.2 & 6.7 \\
\hline
Energy scale     & 3.9 & 13.1 \\
Yield extraction & 3.9 & 2.0  \\
Yield correction & 6.4 & 6.0  \\
\end{tabular} \end{ruledtabular}
\end{table}

The data sets from the two EMCal subsystems, PbSc and PbGl, were 
analyzed separately and combined for the final results. Results from 
the two subsystems were consistent within uncertainties. The 
systematic uncertainty of the combined result is reduced as the major 
systematic uncertainties in the two EMCal subsystems are not 
correlated. For final $\pi^0$ cross section results, BBC-triggered 
events were used for $p_T<3$ GeV/$c$ and high $p_T$ photon triggered 
events in coincidence with the BBC trigger were used for $p_T>3$ 
GeV/$c$.

\subsection{Vernier scan analysis}
\label{label:vernier}

The measured $\pi^0$ cross section was normalized to the integrated 
luminosity for the analyzed data sample ($L$) which was determined 
from the number of BBC-triggered events using an absolute calibration 
of the BBC trigger cross section $\sigma_{BBC}$. The value of 
$\sigma_{BBC}$ is obtained via the van der Meer or Vernier scan 
technique \cite{vdmeer}. This is a crucial part of the absolute cross 
section analysis and is therefore discussed in detail in this section.

In a scan, the transverse widths of the beam overlap $\sigma_x$ and 
$\sigma_y$ were measured by sweeping one beam across the other in 
small steps while monitoring the BBC trigger rate. Then the 
instantaneous machine luminosity of each bunch crossing $L_{machine}$ 
is computed as:
\begin{equation}
L_{machine} = \frac{f_{rev}}
{2 \pi \sigma_x \sigma_y} \cdot N_B \cdot N_Y,
\label{eq:Lmachine}
\end{equation}
where $N_{B}$ and $N_{Y}$ are the bunch intensities of the two beams 
($\sim$$10^{11}$/bunch), $f_{rev}$ is the revolution frequency 
(78~kHz). The BBC trigger cross section $\sigma_{BBC}$ is the ratio of 
the BBC trigger rate when the beams were overlapping maximally 
($R_{max}$) to the effective luminosity $L_{eff}$:
\begin{equation}
\sigma_{BBC} = R_{max} / L_{eff},
\end{equation}
where
\begin{equation}
L_{eff} = L_{machine}
\cdot \epsilon_{\rm vertex},
\label{eq:Leff}
\end{equation}
and $\epsilon_{\rm vertex}$ is the fraction of the number of collisions in 
the PHENIX interaction region (IR) within the BBC trigger vertex cut 
(usually $|z|<30$ cm).

$L_{machine}$ was corrected for the $z$ dependence of the transverse 
beam sizes caused by the beam focusing in the IR (hour-glass effect) 
and for the beam crossing angle. The value of $\epsilon_{\rm vertex}$ was 
extracted from the $z$-vertex distribution of events measured by the 
BBCs and was corrected for the dependence of the BBC trigger 
efficiency on the collision vertex position $z$ along the beam axis. 
These corrections are discussed in more detail below.

\begin{figure}[tb]
\vspace{-0.1in}
\begin{center}
\includegraphics[width=1.0\linewidth]{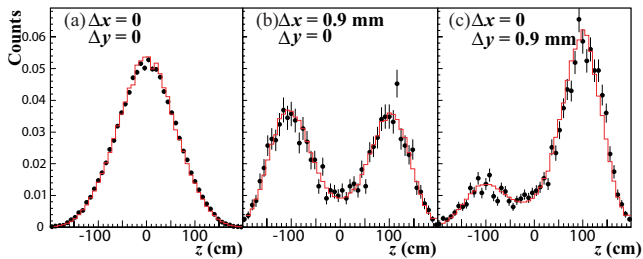}
\end{center}
\caption{\label{fig:vernier} (color online)
Collision $z$-vertex distribution in the PHENIX IR measured 
by ZDCs in a Vernier scan at $\sqrt{s}=200$ GeV (points) and 
calculations from convolution of colliding bunch intensity profiles 
along $z$-axis and including the hour-glass effect for $\beta^*=1$ m, 
for bunches with typical length of 1 m and transverse size of 0.3 mm 
(histograms);
(a) beams are head-on; 
(b) one beam is 0.9 mm displaced relative to the other beam 
in the horizontal direction (illustrates the hour-glass effect) and 
(c) one beam is 0.9 mm displaced relative to the other beam 
in the vertical direction. The calculations include the bunch 
crossing angle with a vertical projection of 0.15 mrad. }
\end{figure}

In $p+p$ collisions at $\sqrt{s}=62.4$ GeV, the BBC trigger efficiency 
vs $z$ shape was estimated from the comparison with a ``detector 
unbiased'' $z$-vertex distribution obtained from the convolution of 
colliding bunch intensity profiles along the $z$-axis as measured by 
Wall Current Monitors (WCMs) \cite{wcm}. The correction factor of 
$0.83 \pm 0.08$ for $\epsilon_{\rm vertex}$ in Eq.~(\ref{eq:Leff}) was 
obtained, resulting in $\epsilon_{\rm vertex}=0.37 \pm 10\%$. This 
approach is confirmed in $p+p$ collisions at $\sqrt{s}=200$~GeV where 
the ZDCs have enough efficiency to measure the $z$ vertex 
distribution. The efficiency of the ZDCs (located at $z=\pm18$ m) does 
not depend on collision vertex position in the PHENIX IR, which was 
distributed with a sigma of 0.5 -- 0.7 m around $z=0$. The vertex 
distribution obtained with the WCMs is well reproduced by the 
measurement with the ZDCs at $\sqrt{s}=200$ GeV 
(Fig.~\ref{fig:vernier}a).

Beam focusing in the IR causes bunch transverse sizes to vary away 
from the nominal collision point ($z=0$) as $\sigma^2(z) = 
\sigma^2(z=0) \times (1+z^2/\beta^{*2})$, where $\beta^{*}$ is the 
value of the betatron amplitude function at the interaction point. 
This is the so-called hour-glass effect.  The product 
$\sigma_x\sigma_y$ in Eq.~(\ref{eq:Lmachine}) should be replaced by an 
effective $\langle \sigma_x \cdot \sigma_y \rangle$, which differs 
from what was measured in a scan (mainly due to the vertex cut 
implemented in BBC trigger). The correction due to this effect for 
Vernier scan data at $\sqrt{s}=62.4$ GeV with a betatron amplitude 
function at the collision point of $\beta^*=3$ m was simulated with 
WCM data and calculated to be $0.93 \pm 0.02$. The applicability of 
our calculational technique is illustrated in Fig.~\ref{fig:vernier} 
with the high statistics Vernier scan data at $\sqrt{s}=200$ GeV.

Figure~\ref{fig:vernier}b and ~\ref{fig:vernier}c shows the 
sensitivity of our data for the transversely displaced beams to the 
hour-glass effect and to the crossing angle between the colliding 
beams, compared with a head-on vertex distribution in 
Fig.~\ref{fig:vernier}a. The two peaks in Fig.~\ref{fig:vernier}b and 
~\ref{fig:vernier}c, caused by the hour-glass effect, show an overlap 
of the diverging colliding beams at large $|z|$ in a particular 
displaced beam setting from a Vernier scan. The obvious asymmetry in 
the two peaks in Fig.~\ref{fig:vernier}c is a result of the non-zero 
crossing angle between colliding bunches. In all Vernier scan 
measurements the crossing angle was found to be less than 0.2 mrad, 
which translates to a negligible correction for $L_{machine}$ at 
$\sqrt{s}=62.4$ GeV, with a typical bunch length of $\sim 1$ m and 
bunch transverse size of $1$ mm.

\begin{figure}[tb]
\includegraphics[width=1.0\linewidth]{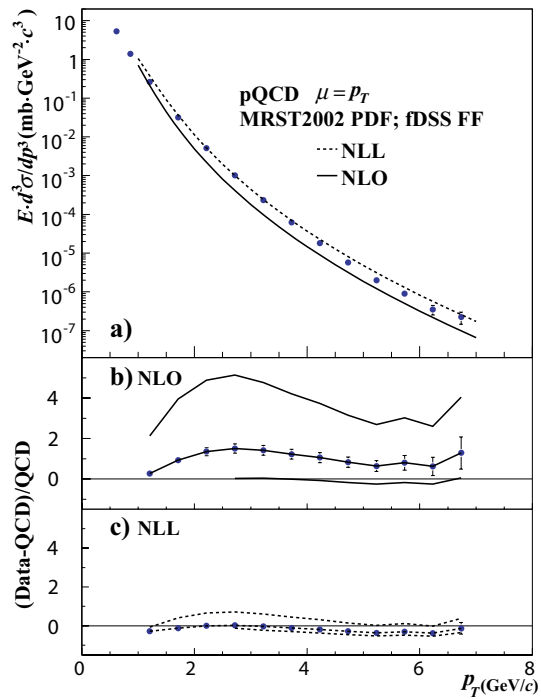}
\caption{\label{fig:cross_nlo}  (color online)
(a) The neutral pion production cross section 
at $\sqrt{s}=62.4$~GeV as a function of $p_{T}$ (circles)
and the results of NLO (solid) and NLL (dashed) pQCD calculations 
for the theory scale $\mu=p_T$.
(b) The relative difference between the data and NLO pQCD calculations 
for the three theory scales $\mu=p_T/2$ (upper line), $p_T$ (middle 
line) and $2p_T$ (lower line); experimental uncertainties (excluding the 
11\% normalization uncertainty) are shown for the $\mu=p_T$ curve.
(c) The same as b) but for NLL pQCD calculations.
}
\end{figure}

\begin{figure}[tb]
\includegraphics[width=1.0\linewidth]{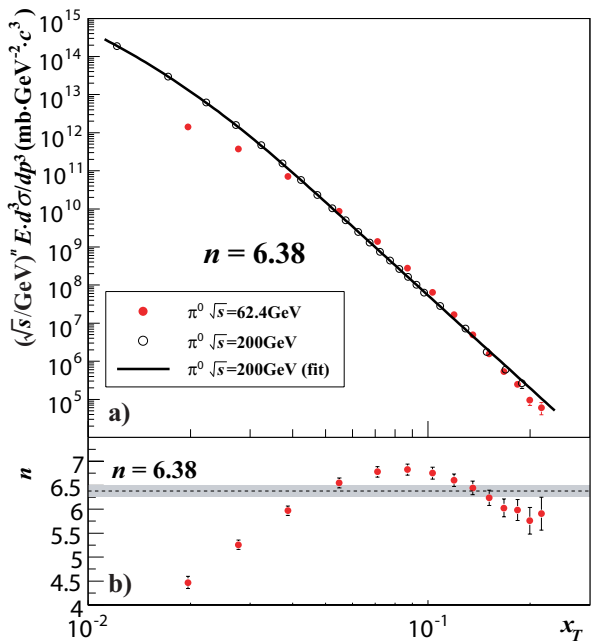}
\caption{\label{fig:cross_xt} (color online)
(a) The neutral pion production cross section 
at $\sqrt{s}=62.4$~GeV and $\sqrt{s}=200$~GeV as a function of $x_{T}$, 
scaled by $(\sqrt{s}/{\rm GeV})^n$ with n=6.38; 
the solid line is a parameterization of $\sqrt{s}=200$~GeV data.
(b) The parameter $n$ in (\ref{eq:xt_scale}) obtained from the ratio 
of invariant cross section at $\sqrt{s}=62.4$~GeV and 
$\sqrt{s}=200$~GeV, at each $x_T$ of $\sqrt{s}=62.4$~GeV data; error 
bars show the statistical and systematic uncertainties of the 
$\sqrt{s}=62.4$~GeV and $\sqrt{s}=200$~GeV data. The shaded band 
reflects the $11\% \oplus 9.7\%$ normalization uncertainty in the 
$\sqrt{s}=62.4$ and 200~GeV cross section measurements, 
correspondingly.
}
\end{figure}

After all the corrections discussed above were applied, our BBC 
trigger cross section in $p+p$ collisions at $\sqrt{s}=62.4$ GeV was 
found to be $\sigma_{BBC} = 13.7$ mb with a systematic uncertainty of 
$\pm 1.5$ mb ($\pm 11\%$), i.e. $\sim 40$\% of the world-average value 
of the inelastic $p+p$ scattering cross section at 
$\sqrt{s}=62.4$~GeV\cite{david_ISR}. Major contributors to the 
systematic uncertainty are $4\%$ from the uncertainty in the 
normalization of bunch intensity measurements and in the calibration 
of the beam position measurements in the Vernier scan, $10\%$ from the 
BBC trigger efficiency correction of $\epsilon_{\rm vertex}$, and $2\%$ 
from the hour-glass correction.

\subsection{$\pi^0$ cross section results and discussion}
\label{label:xsec_results}

Figure~\ref{fig:cross_nlo} presents the inclusive mid-rapidity $\pi^0$ 
invariant production cross section at $\sqrt{s}=62.4$ GeV versus 
$p_T$, from $p_T=0.5$ GeV/$c$ to $p_T=7$ GeV~\cite{data}. An overall 
normalization uncertainty of 11\% due to the uncertainty in absolute 
normalization of the luminosity is not shown. The analyzed data sample 
with $0.76\times 10^{9}$ BBC triggers corresponded to about 55 
nb$^{-1}$ integrated luminosity. The measurements fall within the 
large spread of ISR data~\cite{ISR_n,ISR_pi,david_ISR}.

The data are compared to NLO and NLL pQCD calculations at a theory 
scale $\mu=p_T$, where $\mu$ represents equal factorization, 
renormalization, and fragmentation scales~\cite{nlo_nll}. The NLL 
corrections extend the NLO calculations to include the resummation of 
extra ``threshold'' logarithmic terms which appear in the perturbative 
expansion at not very high energies because the initial partons have 
just enough energy to produce the high $p_T$ parton that fragments 
into a final pion. The MRST2002 parton distribution 
functions~\cite{MRST2002} and the fDSS set of fragmentation 
functions~\cite{fDSS}, which are extracted in NLO, are used in both 
NLO and NLL calculations. We have previously seen that the data are 
well described by NLO pQCD with a scale of $\mu=p_T$ at $\sqrt{s}=200$ 
GeV~\cite{pi0cross_run2,pi0cross_all_run5}. In contrast, NLO 
calculations with the same scale underestimate the $\pi^0$ cross 
section at $\sqrt{s}=62.4$ GeV. At the same time, it is known that NLO 
calculations are not always successful at describing low energy fixed 
target data~\cite{low_energy}, while NLL calculations have been 
successful~\cite{nll}. The NLL calculations have a smaller scale 
dependence and describe our data well with $\mu=p_T$;
however, as noted in~\cite{nlo_nll}, subleading perturbative corrections
to the NLL calculation may be significant. Therefore, the 
results may indicate that $\sqrt{s}=62.4$ GeV is at an intermediate 
energy region where calculations that include threshold logarithm 
effects may describe the data more accurately. Therefore, below we 
show comparisons to both NLO and NLL at a scale of $\mu=p_T$.

General principles of hard scattering, including the principle of 
factorization of the reaction into parton distribution functions for 
the protons, fragmentation functions for the scattered partons and a 
short-distance parton-parton hard scattering cross section, predicted 
a general $x_T$-scaling form for the invariant cross section of 
inclusive particle production near mid-rapidity \cite{xtscaling_orig}:
\begin{equation}
E\frac{d^3\sigma}{dp^3}=\frac{1}{p_T^n}F(x_T)=
\frac{1}{\sqrt{s}^{\tiny{\ }n}}G(x_T)
\label{eq:xt_scale}
\end{equation}
where $x_T=2p_T/\sqrt{s}$, and $F(x_T)$ and $G(x_T)$ are universal 
functions. The parameter $n$ relates to the form of the force-law 
between constituents. For example for QED or Vector Gluon exchange, 
$n=4$ \cite{scale_qed}. Due to higher order effects, the running of 
the coupling constant $\alpha(Q^2)$, the evolution of the parton 
distribution functions and fragmentation functions, and the 
initial-state transverse momentum $k_T$, $n$ is not a constant but is 
a function of $x_T$ and $\sqrt{s}$: $n(x_T,\sqrt{s})$ 
\cite{xtscaling}.

Figure~\ref{fig:cross_xt}a shows the inclusive $\pi^0$ cross section 
scaled by $\sqrt{s}^{\scriptsize{\ }n}$ for $\sqrt{s}=62.4$~GeV and 
200~GeV data \cite{pi0cross_all_run5} as a function of $x_T$, with the 
parameter $n=6.38$, which is a weighted average of $n(x_T)$ for 
$x_T>0.07$ (corresponding to $p_T>2$ GeV/$c$ at $\sqrt{s}=62.4$ GeV). 
The parameter $n(x_T)$ was calculated as 
$\ln(\sigma_{62.4}(x_T)/\sigma_{200}(x_T))/\ln(200/62.4)$ for each 
$x_T$ of $\sqrt{s}=62.4$~GeV data, and $\sigma_{62.4}$ and 
$\sigma_{200}$ are invariant differential cross sections at 
$\sqrt{s}=62.4$~GeV and $\sqrt{s}=200$~GeV, respectively. Cross 
section values for the corresponding $x_T$ at $\sqrt{s}=200$~GeV were 
obtained from parameterization of the measured cross section at 
$\sqrt{s}=200$~GeV: $T(p_T)\frac{A}{(1+p_T/p_0)^m} + 
(1-T(p_T))\frac{B}{p_T^k}$, $T(p_T) = \frac{1}{1+\exp((p_T-t)/w)}$, 
where $t=4.5$~GeV/$c$, $w=0.084$~GeV/$c$, $A=253.8$ 
mb$\cdot$GeV${}^{-2} \cdot c^{3}$, $p_0=1.488$~GeV/$c$, $m=10.82$, 
$B=14.7$~mb$\cdot$GeV${}^{-2+k}\cdot c^{3-k}$, and $k=8.11$. All 
$\sqrt{s}=200$~GeV data points agree with the parameterization curve 
within uncertainties. The parameterization is shown as the solid curve 
in Fig.~\ref{fig:cross_xt}a.

At low $x_T$, where soft physics dominates particle production, 
$n(x_T)$ is supposed to increase with $x_T$ due to the similar 
exponential shapes of the soft part of the invariant cross section 
versus $p_T$ at different $\sqrt{s}$ ($\sim e^{-6p_T}$) 
\cite{scale_qed}. In the hard scattering region $n(x_T)$ is expected 
to decrease with increasing $x_T$, due to stronger scale breaking at 
lower $p_T$. Such behavior of $n(x_T)$ is demonstrated by our data in 
Fig.~\ref{fig:cross_xt}b. A similar drop in the parameter $n$ at 
$x_T\gtrsim0.1$ was observed at ISR energies \cite{ISR_n}. 
Figure~\ref{fig:cross_xt}b also shows the possible transition from 
soft- to hard-scattering regions in $\pi^0$ production at $p_T \sim 
2$~GeV/$c$. A similar conclusion was derived from the shape of the 
$\pi^0$ spectrum at $\sqrt{s}=200$~GeV in \cite{pi0cross_all_run5}. 
This can serve as a basis for applying the pQCD formalism to the 
double helicity asymmetry data with $p_T>2$ GeV/$c$ in order to allow 
access to $\Delta G$.

\section{Inclusive $\pi^0$ Double Helicity Asymmetry}
\label{label:all}

\subsection{$\pi^0$ $A_{LL}$ analysis}
\label{label:all_analysis}
For the 2006 run, each of the two independent RHIC collider rings were 
filled with up to 111 bunches in a 120 bunch pattern, with one of 
four, fill-by-fill alternating predetermined patterns of polarization 
sign for the bunches. Bunch polarization signs in each pattern were 
set in such a way that all four colliding bunch spin combinations 
occurred in sequences of four bunch crossings. That greatly reduced 
the systematic effects in spin asymmetry measurements due to variation 
of detector response versus time and due to possible correlation of 
detector performance with RHIC bunch structure.

To collect data from collisions of longitudinally polarized protons, 
the polarization orientation of the beams was rotated from vertical, 
the stable spin direction in RHIC, to longitudinal at the PHENIX IR 
and then back to vertical after the IR by spin rotators ~\cite{waldo}. 
PHENIX local polarimeters measured the residual transverse component 
of the beam polarizations, using the spin dependence of very forward 
neutron production ~\cite{locpol,pi0all_run3,pi0cross_all_run5} 
observed by the ZDC, and by that means monitored the orientation of 
the beam polarization in the PHENIX IR throughout the run.

The magnitudes of the beam polarizations at RHIC are measured using 
fast carbon target polarimeters~\cite{pol_pC}, normalized to absolute 
polarization measurements by a separate polarized atomic hydrogen jet 
polarimeter~\cite{pol_jet}. The luminosity-weighted beam polarizations 
over 11 RHIC fills used in the $A_{LL}$ analysis were $\langle P 
\rangle$=0.48 for both beams, with 0.035 and 0.045 systematic 
uncertainty for the two RHIC beams, respectively. For the longitudinal 
polarization run period, the residual transverse polarizations of the 
beams were $\langle P_T/P \rangle^B$=0.11$\pm$0.15 and $\langle 
P_T/P\rangle^Y=0.11\pm0.12$ for ``Blue'' and ``Yellow'' RHIC beams, 
respectively. The average transverse component of the product was 
$\langle P_T^B \cdot P_T^Y \rangle / \langle P^B \cdot P^Y \rangle \le 
\langle P_T/P \rangle^B \cdot \langle P_T/P \rangle^Y = 
0.012\pm0.021$; the average of the polarization product over the run 
was $\langle P^B \cdot P^Y \rangle = 0.23$, with a systematic 
uncertainty of $\pm14\%$.

Experimentally, the double helicity asymmetry for $\pi^0$ production 
is determined as:
\begin{equation}
A_{LL}^{\pi^0} = \frac{1}{|P^B \cdot P^Y|} \cdot \frac{N_{++}-R\cdot
N_{+-}}{N_{++} + R \cdot N_{+-}};~~~R=\frac{L_{++}}{L_{+-}},
\label{eq:a_ll}
\end{equation}
where $N_{++}$ and $N_{+-}$ are the number of $\pi^0$'s and $R$ is the 
relative luminosity between bunches with the same and opposite 
helicities. The analysis technique for the $\pi^0$ $A_{LL}$ 
measurements is similar to our analyses of $\sqrt{s}=200$ GeV 
data~\cite{pi0all_run3,pi0all_run4,pi0cross_all_run5}.

Double helicity asymmetry results were obtained from longitudinally 
polarized $p+p$ collisions corresponding to $\sim 40$ nb$^{-1}$ 
integrated luminosity. Due to the limited BBC trigger efficiency for 
high $p_T$ $\pi^0$ events, high $p_T$ photon triggered events without 
the BBC trigger condition requirement were used for the $\pi^0$ 
asymmetry analysis. This led to a slightly increased background in the 
$\pi^0$ reconstruction and additional systematic uncertainty in the 
measurements of the relative luminosity between bunches with different 
helicity states.

The background asymmetry under the $\pi^0$ peak in the two-photon mass 
distribution $A_{LL}^{BG}$ was estimated from the counts outside the 
$\pi^0$ peak, from a 177--217 MeV/$c^2$ range in the two-photon mass 
distribution. Unlike our $\sqrt{s}=200$~GeV data analyses, a lower 
mass range was not used for $A_{LL}^{BG}$ estimations due to cosmic 
background from non-collision events. This background contribution in 
the mass ranges of $\pi^0$ peak and higher mass was negligible 
($<1\%$), and $A_{LL}^{BG}$ was consistent with zero in all $p_T$ 
bins.

Similar to our previous analyses, crossing-by-crossing accumulated 
number of BBC triggers were used for the measurements of the relative 
luminosity between bunches with different spin configuration. The 
uncertainty on the relative luminosity measurements $\delta R$ was 
derived from the comparison between BBC trigger events and other 
trigger events, selecting different physics processes in different 
kinematic ranges. In the $\sqrt{s}=200$~GeV data analysis the 
comparison was done to triggers defined by the coincidence of signals 
from the two ZDCs \cite{pi0all_run3,pi0all_run4,pi0cross_all_run5}. 
Due to the limited efficiency of the ZDC at $\sqrt{s}=62.4$~GeV, the 
comparison in this analysis was performed with the number of events 
which fired simultaneously either of the two BBCs and either of the 
two ZDCs. Only 20\% of the event statistics in this sample is 
contributed by BBC-triggered events, so this sample can be considered 
as essentially independent from the BBC event sample. From this 
comparison the upper limit of $\delta R$ was estimated to be 
$0.6\times10^{-3}$, which for the average beam polarizations of 0.48 
translates to $\delta A_{LL}=1.4\times10^{-3}$, the $p_T$ independent 
uncertainty of the $\pi^0$ double helicity asymmetry results. Single 
beam background $<0.35\%$, as determined by the trigger counts of 
non-colliding bunches and pileup probability of $\alt 0.02\%$, had 
negligible impact on the relative luminosity measurements.

A transverse double spin asymmetry $A_{TT}$, the transverse equivalent 
to Eq.~(\ref{eq:a_ll_sigma}) and (\ref{eq:a_ll}), can contribute to 
$A_{LL}$ through the residual transverse component of the product of 
the beam polarizations discussed above. Similar to 
\cite{pi0cross_all_run5}, $A_{TT}$ was obtained from the sample with 
transverse polarization. The maximal possible $A_{TT}$ effect on 
$A_{LL}$ was determined by $\pm \delta A_{TT}$ from the measured 
$A_{TT}$, which was $<0.15\cdot\delta A_{LL}$ in all $p_T$ bins.

\subsection{$\pi^0$ $A_{LL}$ results and discussion}
\label{label:all_results}

\begin{figure}[tb]
  \includegraphics[width=1.0\linewidth]{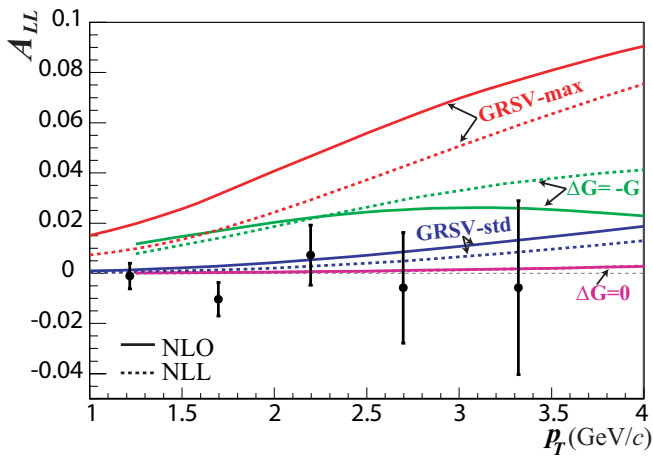}
	\caption{\label{fig:all} (color online)
The double helicity asymmetry for 
	  neutral pion production at $\sqrt{s}=62.4$ GeV 
	  as a function of $p_T$ (GeV/$c$).
	  Error bars are statistical uncertainties, 
	  with the 14\% overall polarization uncertainty not shown; 
	  other experimental systematic uncertainties are negligible. 
	  Four GRSV theoretical calculations based on NLO pQCD 
          (solid curves) and on NLL pQCD (dashed curves) are also shown
	  for comparison with the data (see text for details).
	  Note that the $\Delta G=0$ curves for NLO and NLL overlap.
	}
\end{figure}

Figure~\ref{fig:all} presents the measured double helicity asymmetry 
in $\pi^{0}$ production versus $p_T$~\cite{data}. A scale uncertainty 
of 14\% in $A_{LL}^{\pi^0}$ due to the uncertainty in beam 
polarizations is not shown. The other systematic uncertainties are 
negligible, as discussed above, and checked using a technique to 
randomize the sign of bunch polarization and by varying the $\pi^0$ 
identification criteria~\cite{pi0all_run3}.

Figure~\ref{fig:all} also shows a set of $A_{LL}$ curves from pQCD 
calculations that incorporates different scenarios for gluon 
polarization within the GRSV parameterization of the polarized parton 
distribution functions~\cite{grsv}. GRSV-std corresponds to the best 
fit to inclusive DIS data. The other three scenarios in 
Fig.\ref{fig:all} (GRSV-max, $\Delta G=0$, and $\Delta G=-G$) are 
based on the best fit, but use the functions $\Delta G(x_g) = G(x_g), 
0, -G(x_g)$ at the initial scale for parton evolution ($Q^{2}=0.4$ 
GeV$^{2}$), where $G(x_g)$ is the unpolarized gluon distribution, and 
$\Delta G(x_g)$ is the difference between the distribution of gluons 
with the same and opposite helicity to the parent proton. In 
Fig.~\ref{fig:all}, we compare our asymmetry data with both NLO and 
NLL calculations. The NLL calculations indicate that we have a reduced 
sensitivity to positive $\Delta G$, but the effect is far less 
pronounced than at Fermilab fixed-target energies~\cite{nlo_nll}. 
Similar to our $\sqrt{s}=200$ GeV results 
\cite{pi0all_run3,pi0cross_all_run5}, our $\sqrt{s}=62.4$ GeV $A_{LL}$ 
data do not support a large gluon polarization scenario, such as 
GRSV-max.

\begin{figure}[tb]
  \includegraphics[width=1.0\linewidth]{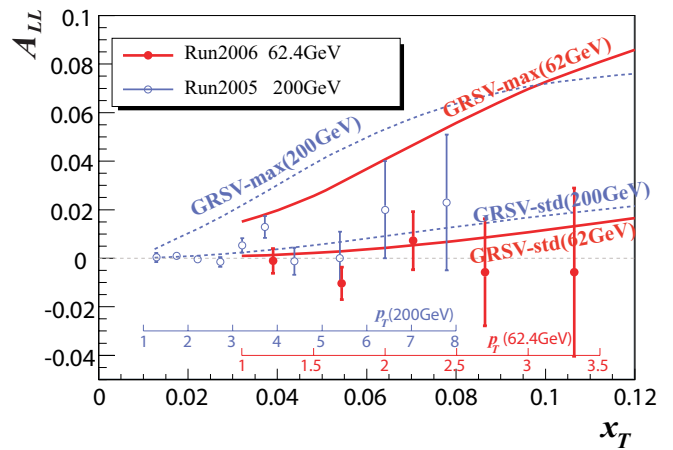}
  \caption{\label{fig:all_xt} (color online)
The double helicity asymmetry for 
    neutral pion production at $\sqrt{s}=62.4$ GeV and 200 GeV
    as a function of $x_T$.
    Error bars are statistical uncertainties, 
    with the 14\% (9.4\%) overall polarization uncertainty
    for $\sqrt{s}=62.4$~GeV (200 GeV) data not shown.
    Two GRSV theoretical calculations based on NLO pQCD
    are also shown for comparison with the data (see text for details.)
  }
\end{figure}

Figure~\ref{fig:all_xt} presents the measured $A_{LL}$ versus $x_T$ in 
$\pi^0$ production overlaid with the results at $\sqrt{s}=200$ GeV 
\cite{pi0cross_all_run5}. Clear statistical improvement can be seen at 
higher $x_T$. For the measured $p_T$ range 2--4 GeV/$c$, the range of 
$x_g$ in each bin is broad and spans the range $x_g=0.06-0.4$, as 
calculated by NLO pQCD~\cite{werner_private}. Thus our data set 
extends our $x_g$ reach of sensitivity to $\Delta G$ and also overlaps 
previous measurements, providing measurements with the same $x_g$ but 
at a different $Q^2$ scale.

\section{SUMMARY}
\label{label:summary}

To summarize, we have presented the unpolarized cross section and 
double helicity asymmetries for $\pi^{0}$ production at mid-rapidity, 
for proton-proton collisions at $\sqrt{s}=62.4$ GeV. The accuracy of 
the cross-section measurements, which fall within the
large spread of ISR data, relies on direct $\pi^0$ two-photon 
decay reconstruction, precise calibration of the photon energy 
measurements, careful study of the trigger performance and accurate 
control of the integrated luminosity of the analyzed data sample.
The results serve as a 
precise baseline for heavy-ion measurements. Comparisons to NLO and 
NLL theoretical calculations indicate that including the effects of 
threshold logarithms may be necessary to more accurately describe the 
cross section at $\sqrt{s}=62.4$ GeV. The $A_{LL}$ results extend the 
sensitivity to the polarized gluon distribution in the proton to 
higher $x_g$ compared to the previous measurements at 
$\sqrt{s}=200$~GeV. A preliminary version of these double helicity 
asymmetry results was already used in a recent global fit of both RHIC 
and polarized DIS data to constrain $\Delta G$ \cite{gl_fit}.

\begin{acknowledgments}


We thank the RHIC Polarimeter Group and
the staff of the Collider-Accelerator and Physics
Departments at Brookhaven National Laboratory and the staff of
the other PHENIX participating institutions for their vital
contributions.  
We thank W.~Vogelsang and M.~Stratmann 
for providing the pQCD calculations 
and for informative discussions. 
We acknowledge support from the 
Office of Nuclear Physics in the
Office of Science of the Department of Energy,
the National Science Foundation, 
a sponsored research grant from Renaissance Technologies LLC, 
Abilene Christian University Research Council, 
Research Foundation of SUNY, 
and Dean of the College of Arts and Sciences, Vanderbilt University 
(U.S.A),
Ministry of Education, Culture, Sports, Science, and Technology
and the Japan Society for the Promotion of Science (Japan),
Conselho Nacional de Desenvolvimento Cient\'{\i}fico e
Tecnol{\'o}gico and Funda\c c{\~a}o de Amparo {\`a} Pesquisa do
Estado de S{\~a}o Paulo (Brazil),
Natural Science Foundation of China (People's Republic of China),
Ministry of Education, Youth and Sports (Czech Republic),
Centre National de la Recherche Scientifique, Commissariat
{\`a} l'{\'E}nergie Atomique, and Institut National de Physique
Nucl{\'e}aire et de Physique des Particules (France),
Ministry of Industry, Science and Tekhnologies,
Bundesministerium f\"ur Bildung und Forschung, Deutscher
Akademischer Austausch Dienst, and Alexander von Humboldt Stiftung (Germany),
Hungarian National Science Fund, OTKA (Hungary), 
Department of Atomic Energy (India), 
Israel Science Foundation (Israel), 
Korea Research Foundation and Korea Science and Engineering Foundation (Korea),
Ministry of Education and Science, Rassia Academy of Sciences,
Federal Agency of Atomic Energy (Russia),
VR and the Wallenberg Foundation (Sweden), 
the U.S. Civilian Research and Development Foundation for the
Independent States of the Former Soviet Union, 
the US-Hungarian Fulbight Foundation for Educational Exchange,
and the US-Israel Binational Science Foundation.

\end{acknowledgments}


\def\PRL{Phys. Rev. Lett.\ }
\def\PRD{{Phys. Rev.}~{\bf D}}
\def\PRC{{Phys. Rev.}~{\bf C}}
\def\PLB{{Phys. Lett.}~{\bf B}}
\def\NIMA{{Nucl. Instrum. Methods}~{\bf A}}


\begin{references}
\bibitem{EMC} J. Ashman {\em et al.} (EMC), Phys. Lett. {\bf B206} 
                   364 (1988), Nucl. Phys. {\bf B328}, 1 (1989).

\bibitem{sDIS}
A.~Airapetian {\em et al.} (HERMES), Phys. Rev. {\bf D75}, 012007 (2007);
V.Yu.~Alexakhin {\em et al.} (COMPASS), Phys. Lett. {\bf B647}, 8 (2007).

\bibitem{rhic_spin} G.~Bunce {\it et al.}, Ann. Rev. Nucl. Part. Sci. 
{\bf 50}, 525 (2000).

\bibitem{pi0cross_run2} S.S.~Adler {\it et al.} (PHENIX), \PRL {\bf 91}, 
241803 (2003).

\bibitem{pi0cross_all_run5} A.~Adare {\it et al.} (PHENIX), \PRD {\bf 
76}, 051106 (2007).

\bibitem{pi0_star} J.~Adams {\it et al.} (STAR), \PRL {\bf 92}, 171801 
(2004).

\bibitem{jet_star} B.I.~Abelev {\it et al.} (STAR), \PRL {\bf 97}, 
252001 (2006).

\bibitem{photon_phenix} S.S.~Adler {\it et al.} (PHENIX), \PRL {\bf 98}, 
012002 (2007).

\bibitem{low_energy} P.~Aurenche {\it et al.}, Eur. Phys. J. {\bf C13}, 
347 (2000).

\bibitem{nll} D.~de~Florian and W.~Vogelsang, \PRD {\bf 71}, 114004 (2005).

\bibitem{nlo_nll} D.~de~Florian, W.~Vogelsang, and F.~Wagner, \PRD {\bf 
76}, 094021 (2007).

\bibitem{ISR_n} A.L.S.~Angelis (CCOR), Phys. Lett. {\bf B79}, 505 (1978);
C.~Kourkoumelis {\it et al.} (R-806), Phys. Lett. {\bf B84}, 271 (1979).

\bibitem{ISR_pi}
A.L.S.~Angelis {\it et al.} (CMOR), Nucl. Phys. {\bf B327}, 541 (1989);
T.~Akesson {\it et al.} (AFS), Sov. J. Nucl. Phys. {\bf 51}, 836 (1990);
C.~Kourkoumelis {\it et al.} (R-806), Z. Phys. {\bf C5}, 95 (1980);
A.~G.~Clark {\it et al.} (CSZ), Phys. Lett. {\bf B74}, 267 (1978);
F.~W.~Busser {\it et al.} (CCRS), Nucl. Phys. {\bf B106}, 1 (1976);
K.~Eggert {\it et al.} (ACHM), Nucl. Phys. {\bf B98}, 49 (1975);
F.~W.~Busser {\it et al.} (CCR), Phys. Lett. {\bf B46}, 471 (1973);
F.~W.~Busser {\it et al.} (CCRS), Phys. Lett. {\bf B55}, 232 (1975);
M.~Banner {\it et al.} (Saclay), Nucl. Phys. {\bf B126}, 61 (1977);
B.~Alper {\it et al.} (BSC), Nucl. Phys. {\bf B100}, 237 (1975).

\bibitem{david_ISR} D.~d'Enterria, J. Phys. {\bf G31}, S491 (2005).

\bibitem{nim_emc}L.~Aphecetche {\it et al.}, \NIMA {\bf 499}, 521 (2003).

\bibitem{nim_bbc} M.~Allen {\it et al.}, \NIMA {\bf 499}, 549 (2003).

\bibitem{nim_zdc} C.~Adler {\it et al.}, \NIMA {\bf 470}, 488 (2001).

\bibitem{pi0all_run3} S.S.~Adler {\it et al.} (PHENIX), \PRL {\bf 93}, 
202002 (2004).

\bibitem{locpol} Y.~Fukao {\it et al.}, \PLB {\bf 650}, 325 (2007).

\bibitem{vdmeer} S.~van der Meer, ISR-PO/68-31, KEK68-64;
A.~Drees and Z.~Xu, PAC-2001-RPAH116 {\it Presented at IEEE Particle
Accelerator Conference (PAC2001), Chicago, Illinois, 18-22 Jun 2001}.

\bibitem{wcm} P.~Cameron et al, Proc. of the Particle Accelerator
Conference, p.2146 (1999). 

\bibitem{pi0all_run4} S.S.~Adler {\it et al.} (PHENIX), \PRD {\bf 73}, 
091102 (2006).

\bibitem{data} Tables of data available at http://www.phenix.bnl.gov
/phenix/WWW/info/data/ppg087\_data.html


\bibitem{MRST2002} A.~D.~Martin {\it et al.}, Eur. Phys. J. C {\bf 35},
325 (2004)

\bibitem{fDSS} D.~de Florian {\it et al.}, Phys. Rev. {\bf D75}, 114010 
(2007).

\bibitem{xtscaling_orig} R.~Blankenbecler, S.J.~Brodsky and J.F~Gunion, 
\PLB {\bf 42}, 461 (1972).

\bibitem{scale_qed} S.M.~Berman, J.D.~Bjorken and J.B~Kogut, 
\PRD {\bf 4}, 3388 (1971).

\bibitem{xtscaling} R.~F.~Cahalan, K.~A.~Geer, J.~B.~Kogut,
and L.~Susskind, \PRD {\bf 11}, 1199 (1975).

\bibitem{waldo} W.W.~MacKay {\it et al.}, Proceedings of 
the 2003 Particle Accelerator Conference, 
edited by J.~Chew, P.~Lucas and S.~Webber, IEEE, p.1697.


\bibitem{pol_pC} O.~Jinnouchi {\it et al.}, 
RHIC/CAD Accelerator Physics Note 171 (2004).

\bibitem{pol_jet} H.~Okada {\it et al.}, \PLB {\bf 638}, 450 (2006).

\bibitem{grsv} B.~J\"ager {\it et al.}, \PRD {\bf 67}, 054005 (2003);
M.~Gl\"uck {\it et al.}, \PRD {\bf 63}, 094005 (2001).

\bibitem{werner_private} M.~Stratmann and W.~Vogelsang, 
hep-ph/0702083 = J. Phys. Conf. Ser. {\bf 69}, 012035 (2007); 
W.~Vogelsang, private communication.

\bibitem{gl_fit} D.~de Florian {\it et al.}, Phys. Rev. Lett. {\bf 101}, 
072001 (2008).

\bibitem{E704} D.L. Adams {\it et al.}, Phys. Lett. B {\bf 261}, 197 (1991).


\end{references}
\end{document}